# New-generation Silicon photonics beyond the singlemode regime

Long Zhang,[1] Shihan Hong,[1] Yi Wang,[1] Hao Yan,[1] Yiwei Xie,[1] Tangnan Chen,[1] Ming Zhang,[1] Zejie Yu,[1] Yaocheng Shi,[1] Liu Liu,[1] and Daoxin Dai,[1,2]

[1]*State Key Laboratory for Modern Optical Instrumentation, Center for Optical & Electromagnetic Research, College of Optical Science and Engineering, International Research Center for Advanced Photonics, Zhejiang University, Zijingang Campus, Hangzhou 310058, China.*
[2]*Ningbo Research Institute, Zhejiang University, Ningbo 315100, China.*
*Corresponding author: dxdai@zju.edu.cn*

The singlemode condition is one of the most important design rules for optical waveguides in guided-wave optics. The reason following the singlemode condition is that higher-order modes might be excited and thus introduce some undesired mode-mismatching loss as well as inter-mode crosstalk when light propagates along an optical waveguide beyond the singlemode regime. As a result, multimode photonic waveguides are usually not allowed. In this paper, we propose the concept of silicon photonics beyond the singlemode regime, developed with low-loss and low-crosstalk light propagation in multimode photonic waveguides with broadened silicon cores. In particular, silicon photonic waveguides with a broadened core region as wide as 2-3 μm have shown an ultra-low-loss of ~0.1 dB/cm for the fundamental mode even without any special fabrication process. A micro-racetrack resonator fabricated with standard 220-nm-SOI (silicon-on-insulator) MPW-foundry processes shows a record intrinsic Q-factor as high as $1.02\times10^7$ for the first time, corresponding to ultra-low waveguide propagation loss of only 0.065 dB/cm, which is ~20 times less than that of regular 450-nm-wide SOI strip waveguides. A high-performance microwave photonic filter on silicon is then realized with an ultra-narrow 3-dB bandwidth of 20.6 MHz as well as a tuning range of ~20 GHz for the first time. An on-chip 100-cm-long delayline is also demonstrated by using the present broadened SOI photonic waveguides with compact 90° Euler-curve bends, and the measured propagation loss is ~0.14 dB/cm, which is higher than the loss estimated for the micro-racetrack resonators due to some defects in chip scale. The proposed concept of silicon photonics beyond the singlemode regime helps solve the issue of high propagation loss and also significantly reduces the random phase errors of light due to the random variations of waveguide dimensions. In particularity it enables silicon photonic devices with enhanced performances, which paves the way for new-generation silicon photonics realizing the large-scale photonic integration. The concept of silicon photonics beyond the singlemode regime can be extended further for any other material platforms, such as silicon nitride and lithium niobite. This brings numerous new opportunities for various applications such as nonlinear photonics, large-scale photonic integration, quantum photonics, microwave photonics, etc.

## INTRODUCTION

Silicon photonics is one of the most widely explored platforms for photonic integrated circuits (PICs) because of its compatibility with the complementary metal-oxide-semiconductor (CMOS) process. Various passive and active silicon photonic devices have been demonstrated with impressive performances in the past two decades[1-4]. The silicon photonic devices usually have very compact footprints, benefitting from the ultra-high refractive-index-contrast (Δ) between silicon and silicon oxide. Therefore, it is possible to monolithically integrate more functional photonic devices on a silicon-on-insulator (SOI) platform to realize large-scale PICs, which has been desired for a long time since integrated photonics was born in 1969[5]. In particular, one should realize that passive photonic elements often account for a large proportion in large-scale PICs and thus high-performance passive photonic devices are playing an indispensable role in many on-chip photonic systems. For passive photonic devices, one of the most important keys is achieving photonic waveguides with ultra-low propagation losses and small bending radii, which enables long on-chip optical delaylines with low losses as well as optical resonators with ultra-high Q-factors, as required for communication network filters and multiplexers[6], optical gyroscope rotational velocity sensors[7], optical buffers[8], and true-time-delay antenna beam-forming and steering networks[9].

As it is well known, optical waveguides are usually required to be singlemode according to the singlemode condition in order to avoid the excess loss and the crosstalk due to the excitation and interference of higher-order modes. For silicon photonic waveguides with a 220-nm-thick top-silicon core-layer, the core region is usually narrower than 450 nm, in which case the singlemode condition is satisfied for TE polarization. For the popular 450-nm-wide singlemode silicon photonic waveguides with an ultra-high Δ, the propagation loss is mainly from the scattering at rough sidewalls and is usually much higher than those low-Δ photonic waveguides. Great efforts have been made to reduce the loss of silicon photonic waveguides by e.g. improving the fabrication processes since silicon photonics was proposed in 1980s[10]. Currently, the propagation loss of a singlemode silicon photonic waveguides with a 450×220 nm$^2$ core region is usually about 1-2 dB/cm. This is sufficiently low for a single compact silicon photonic device which usually has a propagation distance less than 1000 μm. However, for silicon micro-resonators, the Q-factor is usually about only $10^4$-$10^5$, which might be too low to satisfy the requirements for some applications like ultra-narrow-band microwave photonic filters[11]. Furthermore, for large-scale silicon PICs, the total propagation distance might be as long as tens of centimeters and thus high propagation loss of 1-2 dB/cm in regular silicon photonic waveguides is still a big barrier for the development of next-generation silicon photonics.

In order to reduce the propagation loss, some special fabrication processes have been introduced for silicon photonics to smoothen the waveguide surfaces, such as thermal reflow[12,13], chemical mechanical planarization (CMP)[14], multiple exposure techniques[15,16], etc. For example, a post-exposure bake has been employed to reflow the photoresist, in which way the fabricated silicon photonic waveguide has a propagation loss of 0.76 dB/cm and a suspended microring resonator (MRR) with a Q-factor of $9.2\times10^5$ was demonstrated[17]. Another special silicon photonic waveguide with a propagation loss as low as 0.3 dB/cm was realized with an etchless process enabled by selective thermal oxidation, in which way the waveguide sidewall is very smooth with a roughness of 0.3 nm[18]. However, one should note that these special fabrication techniques are a non-standard and incompatible with those standard processes in foundries. Moreover, these fabrication approaches are not available generally for various material platforms. Therefore, it is still very desired to achieve low-loss photonic waveguides by using regular standard fabrication processes, in which case the roughness of the sidewalls and top/bottom surfaces are usually about 2~3 nm and 0.3 nm[19]. A potential way is to weaken the optical field intensity at the boundaries of a photonic waveguide, so that light scattering by the rough surface is reduced greatly.

Previously, the introduction of an ultrathin core layer had been proposed as an effective approach to achieve very low-loss photonic waveguides[20-22]. For example, 50-nm-thick SiN optical waveguides have been realized with an ultra-low loss of <0.1 dB/m[20]. Similarly, a 60-nm-thick SOI nanophotonic waveguide has been demonstrated with a loss of about 0.61 dB/cm[22]. However, for a photonic waveguide with an ultrathin core layer, the weak optical confinement makes it not work for sharp bends and thus prevents to achieve high integration density. Furthermore, the ultrathin silicon core-layer is also incompatible with the mature and popular 220-nm-thick SOI platform and thus is disadvantageous to further develop large-scale photonic integration. Later, an MRR with a Q-factor of $2.2 \times 10^7$ was obtained by using 1.22-μm-thick SOI ridge waveguides which were fabricated through CMOS-incompatible processes including photoresist reflowing as well as oxidation smoothing[12]. However, the bending radius is as huge as 2450 μm due to the weak confinement ability, which strongly prevents to realize high integration density. Besides, such special fabrication processes usually are not available in foundries due to the incompatibility, and it would be difficult to control the waveguide dimensions precisely due to the oxidation. Recently, the method of using a straight multimode ridge waveguide was proposed for realizing low propagation loss of 0.3 dB/cm[23], as well as an MRR with a Q-factor of $1.1 \times 10^6$. For the demonstrated high-Q MRR, singlemode bent waveguides were introduced and connected to the multimode ridge waveguides through *long* adiabatic waveguide tapers, in which way the effective bending radius is still quite large and the waveguide tapers as well as the singlemode bent waveguides introduce notable mode-conversion loss and scattering loss from their sidewalls. As the cross section of the ridge waveguide are further increased, a lower waveguide loss of 0.026 dB/cm was obtained[24], which is consistent with our following theoretical analysis. More recently, we successfully demonstrated a compact foundry-fabricated MRR with an ultrahigh Q-factor of $2.3 \times 10^6$ by implementing 1.6-μm-wide uniform multimode strip waveguides with sharp Euler-curve bends on a standard 220-nm-SOI platform[25].

In this paper, a silicon photonic waveguide enabling an ultra-low-loss of about 0.1 dB/cm is proposed with a core region broadened to beyond the singlemode regime according to the comprehensive analysis of the light-matter interaction behavior, even without any special fabrication process. The micro-racetrack resonator fabricated with standard 220-nm-SOI MPW-foundry processes has a compact footprint of ~0.17 × 0.9 mm² with the help of Euler-bends and shows a record intrinsic Q-factor as high as $1.02 \times 10^7$ for the first time. It indicates that the waveguide propagation loss is only 0.065 dB/cm, which is about 20 times less than that of regular SOI strip waveguides. The present ultra-high-Q micro-racetrack resonator is then applied to work as a high-performance microwave filter with an ultra-narrow 3-dB bandwidth of 20.6 MHz as well as a tuning range of ~20 GHz for the first time. We also demonstrate a 100-cm-long delay line based on our broadened SOI nanophotonic waveguides with compact 90° Euler-curve bends. The delayline waveguide is spiraled around the edges of the chip and the measured propagation loss is ~0.14 dB/cm, which is higher than the loss estimated for the micro-racetrack resonators due to some defects in chip scale but still well verifies the potential for achieving ultra-low-loss wafer-scale light propagation. The proposed concept of introducing silicon photonic waveguide with a broadened core region helps solve the issue of high propagation loss, which also significantly reduces the random phase errors of propagating light due to the random variations of waveguide dimensions. Therefore, it enables silicon photonics with enhanced performances and thus is very attractive for next-generation large-scale silicon photonic integration. Furthermore, such the concept of enhanced silicon photonics beyond the singlemode regime can be extended for any other material platforms, such as silicon nitride and lithium niobite.

## RESULTS
**Ultra-high-Q micro-racetrack resonator on an SOI platform.**

Figures 1a and 1b show a three-dimensional (3D) view and top view of the proposed ultra-high-Q micro-racetrack resonator. The resonator is mainly composed of two multimode straight waveguides (MSWs) connected with two 180° modified Euler multimode waveguide bends (MWBs). Figure 1c is the cross-sectional view of the silicon photonic waveguide, whose core width is broadened to be beyond the singlemode regime in order to weaken the light-sidewall interaction and thus greatly reduce light scattering loss at the sidewalls. Here a comprehensive analysis for the scattering loss is given by using a 3D volume-current method[26], in which the radiation loss due to the scattering from the inhomogeneous refractive index at rough surfaces is modeled as an equivalent polarization volume current density (see Supplementary Note 1 for more details). For MPW foundry processes, the roughness for the sidewalls and the top/bottom surfaces are about 3 nm and 0.3 nm[19]. In this case, for a regular singlemode SOI photonic waveguide with a 450-nm-wide and 220-nm-thick core region, the scattering loss from the sidewalls is about 2 dB/cm, which is the dominant compared to the much lower scattering loss of 0.1 dB/cm from the top/bottom surfaces. On the other hand, the scattering loss from the sidewalls can be reduced greatly to be lower than 0.1 dB/cm and even 0.01 dB/cm when the core width is increased to be larger than 1.3 μm and 3 μm, respectively (see Supplementary Fig. 2a), while the surface loss remains constant. As a result, the main source of the interface scattering loss turns to the top/bottom surfaces instead of the sidewalls when the waveguide is wide enough. As shown in Fig. 1d, the total scattering loss from the sidewalls and the top/bottom surfaces can be as low as 0.1 dB/cm when choosing $w_{co}$ = 3 μm, which is far beyond the singlemode regime, and the total scattering loss can be further reduced with a smoother waveguide interface. As a result, the design for the multimode waveguide bends should be careful in order to avoid the excitation of higher-order modes. Otherwise, there will be some significant mode-mismatch loss and inter-mode crosstalk when light propagates along silicon photonic circuits with straight-bent waveguide junctions. Usually the radius for multimode waveguide bends is required to be as large as several hundreds of microns[27]. In order to be compact for achieving high-density as well as large free-spectral ranges (FSRs) for resonators, here MWBs based on a modified Euler curve is introduced, whose curvature radius is varied from the maximum $R_{max}$ to the minimum $R_{min}$, as defined by

$$\frac{d\theta}{dL} = \frac{1}{R} = \frac{L}{A^2} + \frac{1}{R_{max}}, \tag{1}$$

where $L$ is the curve length, $A$ is a constant given by $A = [l_{total}/(1/R_{min} - 1/R_{max})]^{1/2}$, in which $l_{total}$ is the total length of the Euler-curve bend. As shown in Fig. 1b, the 180°-bend is realized by a pair of 90° Euler-bends. Basically, the maximum $R_{max}$ should be large enough to ensure the negligible mode mismatch at the MSW-MWB junction, while the minimum $R_{min}$ should be chosen to be adiabatic and minimize the bending loss as well as the footprint (see more details in Supplementary Note 2).

Figure 1e shows the calculated mode-excitation ratios (MERs) at the junction between the multimode straight section and the multimode bent section as the radius $R_{max}$ varied when the $TE_0$ mode is launched. It can be seen that the mode-mismatching loss is less than 0.0004 dB, and the maximal MER corresponding to the $TE_1$ mode is less than −30 dB when $R_{max}$ >1500 μm. Figure 1f shows the calculated light transmissions in the silicon photonic circuit consisting of an input multimode straight section, a 180°-Euler multimode waveguide bend (MWB), and an output multimode straight section when $R_{min}$ = 60 μm. It can be seen that the inter-mode crosstalk is below −25 dB in a broadband wavelength-band. As a result, we choose $R_{max}$ = 1500 μm and $R_{min}$ = 60 μm in this paper, and the simulated light propagation in the designed 180°-Euler MWB and the mode profiles at the input and output ports are shown in Fig. 1g. As shown in this figure, no multimode interference happens almost while the optical field is well confined in the silicon-core region and has little interaction with the sidewalls. For the coupling region of the micro-racetrack resonator, the gap width $w_{gap}$ and the length $l_c$ are optimized so that sufficient coupling is achieved with little higher-order mode excitation. In the present design, the power coupling ratio for the designed directional coupler is about 0.002 by choosing $w_{gap}$ = 0.24 μm and the length $l_c$ = 600 μm (see the details in Supplementary Note 2).

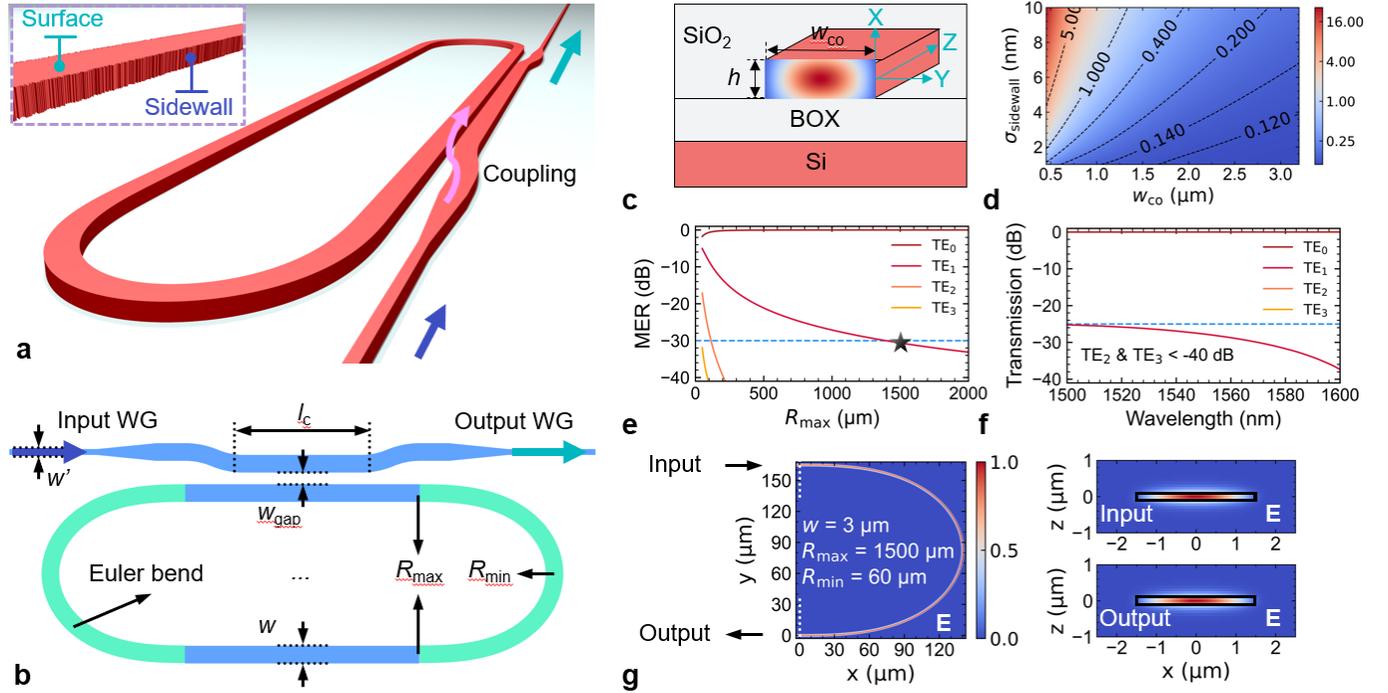

**Fig. 1.** Schematic configurations of the proposed ultra-high-Q micro-racetrack resonator. (a) 3D view; inset: the waveguide sidewall; (b) Top view; (c) Cross section of an SOI photonic waveguide as well as the field distribution of the fundamental mode; (d) Calculated total scattering loss as the waveguide core width $w_{co}$ and the sidewall roughness $\sigma_{sidewall}$ vary. Here assume that the top/bottom-surface roughness $\sigma_{surface}$ = 0.3 nm; (e) Calculated mode-excitation ratios (MERs) at the junction between the multimode straight section and the multimode bent section as the radius $R_{max}$ varied when the $TE_0$ mode is launched; (f) Calculated light transmissions in the silicon photonic circuit consisting of an input multimode straight section, a 180°-Euler MWB, and an output multimode straight section when $R_{min}$ = 60 μm; (g) Simulated light propagation in the designed 180°-Euler MWB and the mode profile at the input and output ports.

**Fabrication and Characterization of the ultra-high-Q micro-racetrack resonators.**

Figure 2a shows the microscope image of the fabricated ultra-high-Q racetrack resonator with a compact size of 0.17×0.9 mm². The scanning electron microscopic (SEM) images for the modified Euler-bend, the coupling region, and the grating coupler are shown in Figs. 2b, 2c and 2d, respectively. Figure 2e shows the measured spectral response at the through port of the resonator, which was normalized with respect to the transmission of an adjacent straight waveguide on the same chip. Here light was launched form a tunable laser (Agilent 81940A) with a minimal step size of 1 pm. The polarization of light from the laser was adjusted by a polarization controller to maximizing the coupling efficiency between the fiber and the grating couple. Finally, the transmission at the output port was monitored by a power meter (Agilent 81618A). Unfortunately, the wavelength resolution for the tunable laser is still limited for the characterizing the present ultra-high-Q resonator. Nevertheless, it can be seen that the racetrack resonator with 620-μm-long MSWs has an FSR of about 0.325 nm, while there are not notable peaks for the higher-order modes, as predicted by our theoretical analysis.

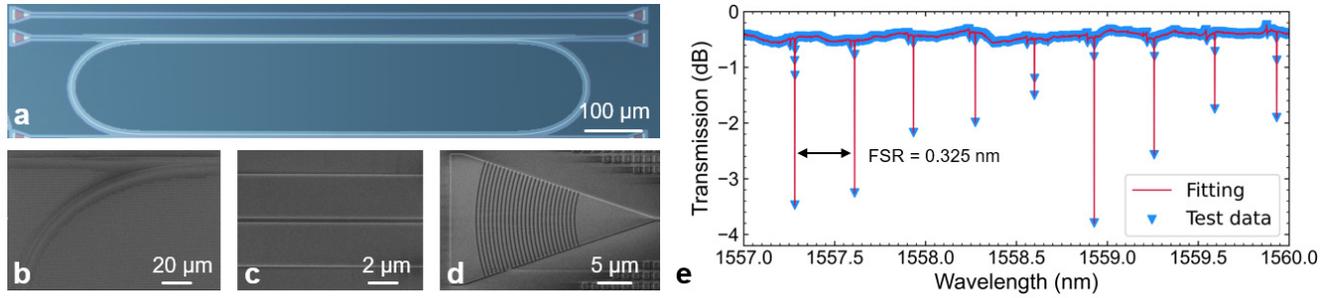

**Fig. 2.** (a) Microscope images of the fabricated ultra-high-Q racetrack resonator; (b) SEM images of the modified Euler bend; (c) Coupling region; (d) Grating couplers; (e) Measured spectral response at the output port.

**Demonstration of ultra-narrow-band microwave photonics filters.**

Figure 3a shows the experimental setup for characterizing the Q factor of the fabricated ultra-high-Q racetrack resonator, which is also used to realize an integrated microwave photonic filter (IMPF) with a very narrow bandwidth. Here a tunable laser was used as the input signal (inset i) and the signal after the PM is in the format of optical double sideband modulation (ODSB) with two sidebands which have a phase difference of π (inset ii). This signal is then sent to the resonator and collected by a PD (xpdv2120ra) which converts an optical signal to a RF signal (inset iii-iv). The RF output is finally monitored by a vector network analyzer (VNA) (Rohde & Schwarz ZVA40), generating a swept RF signal for a PM (inset v). When the upper sideband of the ODSB signal is filtered out by the resonator, the optical carrier and the lower sideband would beat with each other, and then a microwave signal can be generated through a PD. In this case, the microwave signal can well reflect the resonator response and thus one can precisely measure the Q factor (see the details in Supplementary Note 3). Meanwhile, the ultra-high-Q resonator can be used as a RF filter with a very narrow bandwidth. Figure 3b shows the resonance peak of the present ultra-high-Q racetrack resonator, and the measured data can be fit very well by Lorentzian transmissions (see the red curve). It is shown that a full width at half maximum (FWHM) of 20.6 MHz was demonstrated with a resolution of 2 MHz, indicating that the $Q_{load}$ of the racetrack resonator is $0.94 \times 10^7$. The estimated propagation loss for the ultra-high-Q racetrack resonator is as low as about 0.065 dB/cm, corresponding to a record intrinsic Q-factor $Q_i$ more than 10 million (i.e., $1.02 \times 10^7$), which is the best result realized with standard foundry processes.

In addition, here the racetrack-resonator with different core-widths are also demonstrated, and they have similar FSRs of 0.325 nm, as predicted theoretically. For those resonators with different core widths of $w_{co}$ = 0.45 μm, 1.6 μm, and 2 μm, the estimated Q-factors are respectively $1.93 \times 10^5$, $2.73 \times 10^6$ and $3.95 \times 10^6$, corresponding to the propagation losses of 1.6 dB/cm, 0.15 dB/cm, and 0.1 dB/cm, respectively, which agrees well with the theoretical results from the 3D volume current model with $\sigma_{sidewall}$ = 2.6 nm and $\sigma_{surface}$ = 0.24 nm, as shown in Fig. 3c.

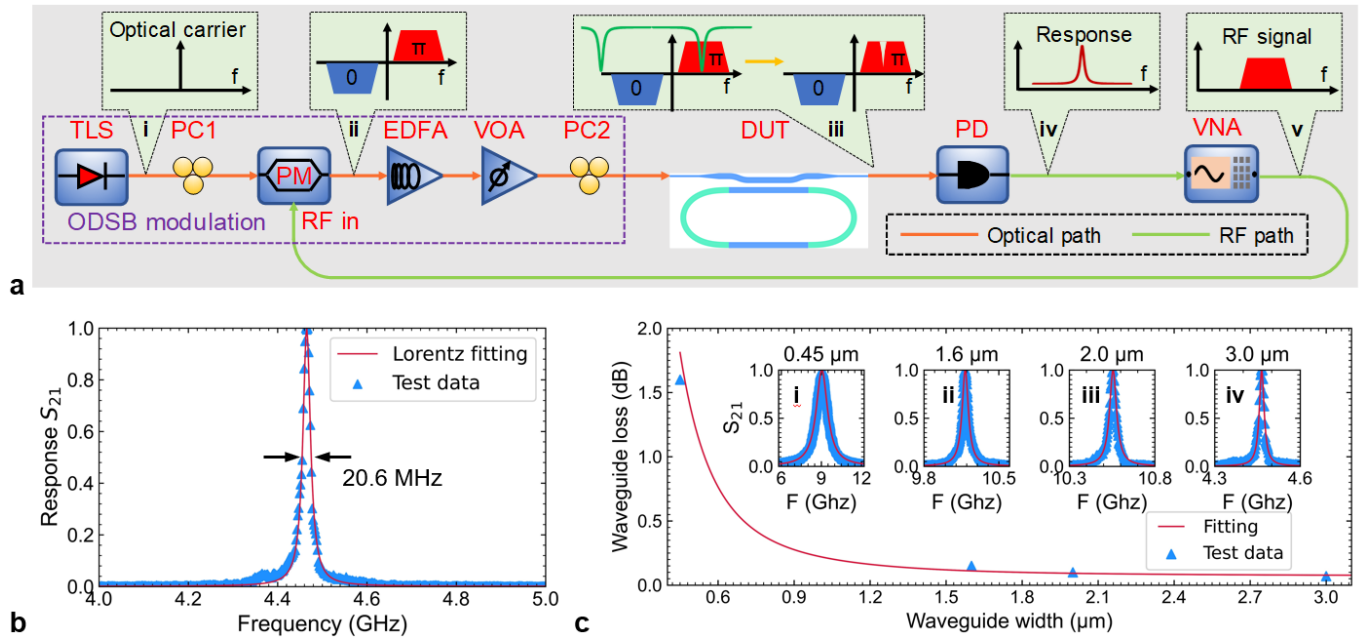

**Fig. 3.** (a) Experimental setup for characterizing the Q factor of the ultra-high-Q racetrack resonator, working as a tunable bandpass IMPF. The insets are: (i) Optical spectra of a high-power laser; (ii) Optical spectrum of the ODSM signal after the PM; (iii) Signal filtered by the resonator; (iv) Electrical spectrum response after the PD; (v) Swept RF single output from the VNA to the PM. Here TL: tunable laser source; PC: polarization controller; EDFA: erbium-doped fiber amplifier; VOA: variable optical attenuator; DUT: device under test; PM: phase modulator; PD: photodetector; VNA: vector network analyzer; (b) Measured result of the resonator with $w_{co}$ = 3 μm by using the microwave method with the Lorentzian-curve fitting; (c) The measured waveguide loss compared with the theoretical prediction. The insets are: the measured resonance peaks of the resonators with different core-widths of $w_{co}$ = 0.45 μm (i), 1.6 μm (ii), 2.0 μm (iii), and 3.0 μm (iv).

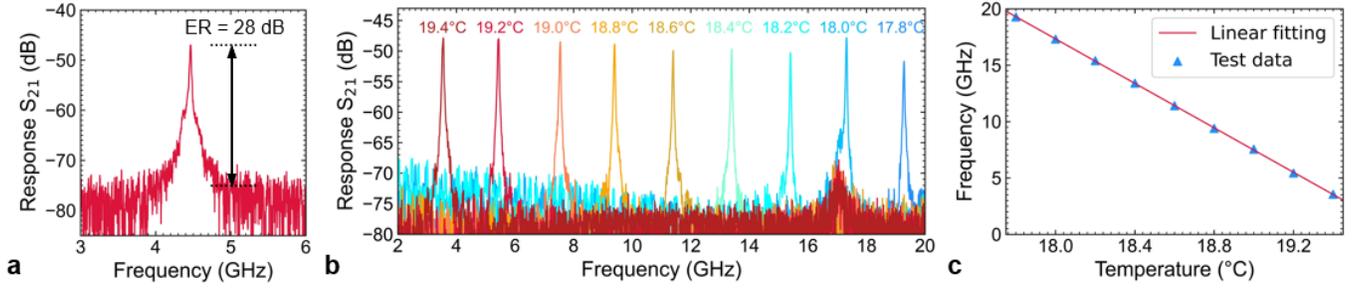

**Fig. 4.** (a) Measured frequency response of the IMPF. (b) Experimental results of the IMPF, demonstrating the tanable central frequency from 3.4 GHz to 19.3 GHz. (c) The measured central frequency as the temperature varies.

With the setup shown in Fig. 3a, we experimentally demonstrate an ultra-narrow band-IMPF based on the proposed ultra-high-Q racetrack resonator. In the experiment, the resonance wavelength of the resonator is controlled precisely by controlling the chip temperature. Figure 4a shows the measured frequency response of the IMPF when its central frequency is 4.45 GHz. The measured extinction ratio (ER) is as high as 28 dB and the 3-dB bandwidth is as small as 20.6 MHz, which is the smallest one among various silicon photonic filters. Figure 4b and 4c shows the measured frequency responses of the IMPF with varied temperature. And the central frequency of the present bandpass IMPF could be tuned from 3.4 GHz to 19.3 GHz by increasing the temperature from 17.8°C to 19.4°C. It can be seen that the proposed IMPF can be tuned very well with ultra-narrowband peaks and flexible tunability.

**100-cm-long low-loss silicon photonic delayline.**
Here, we proposed and demonstrated a 100-cm-long silicon photonic delayline that is laid around the chip with a footprint of 15×3 mm², as shown in Figure 5a. It is composed of 3-μm-wide straight sections connected by 90°-Euler MWBs locating at the corner of the chip. The 3-μm-wide Euler MWBs designed with $R_{max}$ = 1500 μm and $R_{min}$ = 110 μm are utilized in order to avoid the excitation of higher-order modes. With this design, the excess loss at the junction between the multimode straight- and bent-sections is as low as ~0.001 dB and the inter-mode crosstalk is below −30 dB in the wavelength-band around 1550 nm. The width of the gaps between the neighbored waveguides in the spiral is chosen as $w_{gap}$ = 1.06 μm, which is large enough to avoid the crosstalk due to the evanescent coupling between the neighbored waveguide (see more details in Supplementary Note 4). Besides, in order to remove any higher-order mode possibly launched from the input end and filter out any residual higher-order mode at the output end, we introduce short singlemode silicon photonic waveguides at the input and output ends, and 100-μm-long adiabatic tapers are used to connect the singlemode input/output sections and the broadened sections. The designed 100-cm-long silicon photonic delayline was fabricated by the MPW foundry process, as shown in Fig. 5b and 5c. Figures 5d and 5e show the enlarged SEM images of the end-face as well as the sidewalls, verifying that the fabrication is high-quality.

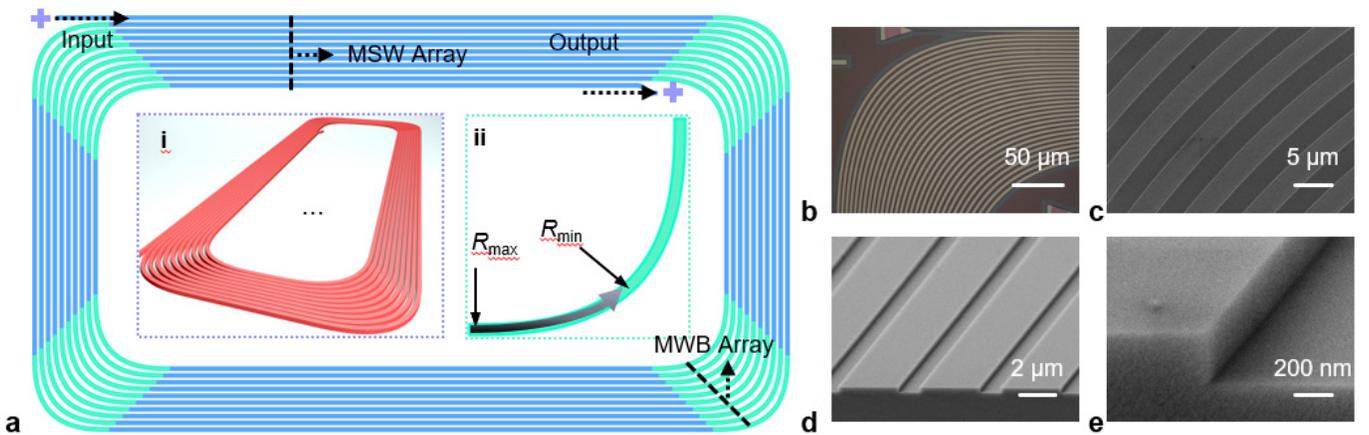

**Fig. 5.** (a) Schematic configurations of the proposed silicon photonic delayline. The insets are the 3D view (i) and the top view of a 90° Euler-bend (ii); Microscope image (b) and SEM image (c) of the Euler-bends of the fabricated silicon photonic delayline; (d) SEM images of the straight sections; (e) Enlarged SEM images of the end face as well as the sidewalls.

In order to characterize the loss of the fabricated silicon photonic delayline, the setup with an amplified spontaneous emission (ASE) source and an optical spectrum analyzer (OSA) was used. The measured transmissions were normalized with respect to the transmission of a 300-μm-long straight waveguide on the same chip, as shown in Figure 6a. According to the measurement results, the propagation loss is estimated to be about 0.12 dB/cm. Figure 6b shows the propagation losses measured from more than 10 chips on the same wafer. It can be seen that the average propagation loss is 0.14 ± 0.025 dB/cm. In order to give a comparison, a 10-cm-long and 0.45-μm-wide silicon photonic delayline was also fabricated on the same chip and the measured results are shown in Fig. 6c and 6d. It can be seen that the average propagation loss is estimated to be 1.4 dB/cm, which is ten times higher than that of the proposed 3-μm-wide silicon photonic delayline. It should be noted that the propagation loss measured from the 100-cm-long silicon photonic delayline is slightly higher than that estimated from the ultra-high-Q resonator. The reason might be that the long delayline has more chances to be with some random defects than a compact resonator and captures attenuation variability caused by wafer-scale variation of the fabrication process[28].

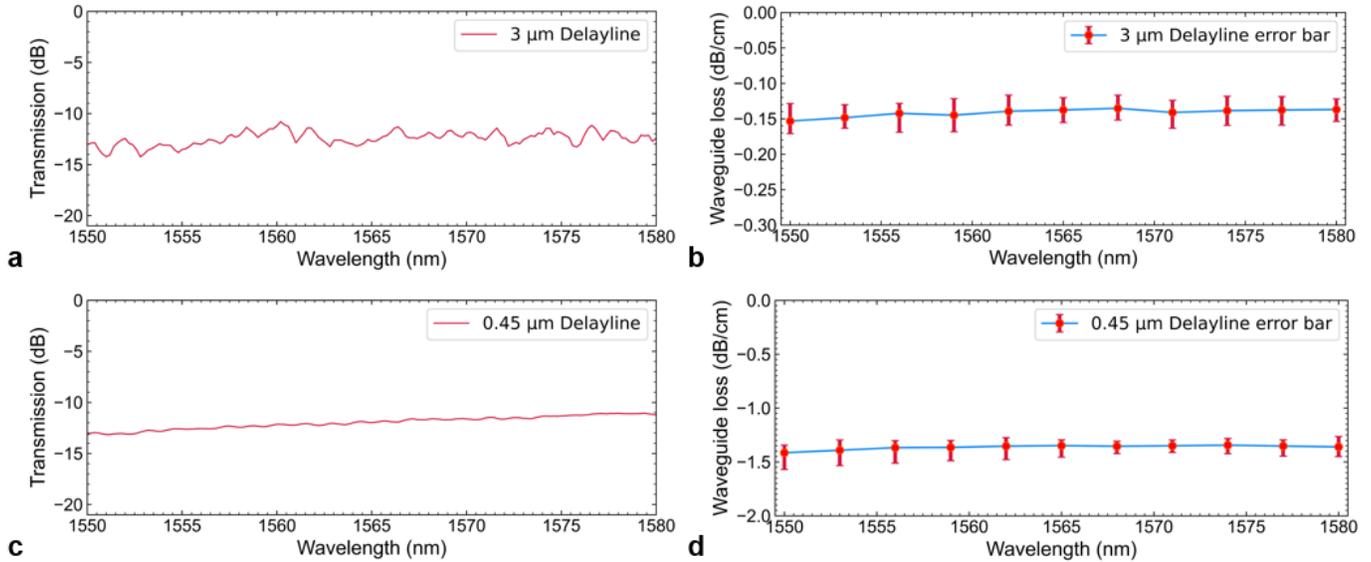

**Fig. 6.** (a) Measured transmissions of the 100-cm-long and **3-μm**-wide silicon photonic delayline on a single chip; (b) Measured propagation losses of the 3-μm-wide silicon photonic delaylines on more than 10 chips; (c) Measured transmissions of the 10-cm-long and **0.45**-μm-wide silicon photonic delayline on a single chip; (d) Measured propagation loss of the 0.45-μm-wide silicon photonic delayline on more than 10 chips.

## DISCUSSION AND CONCLUSION

Figures 7a and 7b show the summary of the recent results of ultra-high-Q silicon photonic resonators and IMPFs, respectively. Among the reported ultra-high-Q resonator on a popular ~220-nm-thick (e.g., 200-260 nm) SOI platform[13,23,25,29-36] (see Supplementary Note 5 for more details), the present racetrack resonator has the record high intrinsic Q-factor of $1.02\times10^7$ even with a regular standard one-step MPW foundry process. In contrast, many high-Q silicon photonic resonators reported previously were often realized with some *special* fabrication processes, such as the photoresist reflowing together with the thermal oxidation for smoothing a ridge waveguide of a large cross-section, which is disadvantageous for further developing large-scale silicon PICs. Meanwhile, the present ultra-high-Q racetrack resonator has a fairly large FSR, benefiting from the adoption of compact 180°-Euler MWBs. This is very useful for realizing microwave photonic filters with relatively large tuning ranges. As shown in Fig. 7b, the present ultra-high-Q resonator provides an option to realize an impressive IMPF with an ultra-narrow-band of 20.6 MHz as well as a tuning range of 20 GHz, which is the best one on a chip reported until now[29,31,32,37-44], even comparable to those filters based on simulated Brillouin scattering[45-47]. Furthermore, the proposed IMPF is well-balanced to satisfy the requirements of ultra-narrowband peak, large operating bandwidth, flexible tunability, compact footprint as well as the CMOS-compatibility. Besides, the present silicon photonic delayline is the longest (as long as 100 cm) and has the lowest propagation loss of ~0.1 dB/cm for uniform *strip* SOI photonic waveguide[18,24,48-54] (see Supplementary Note 5 for more details).

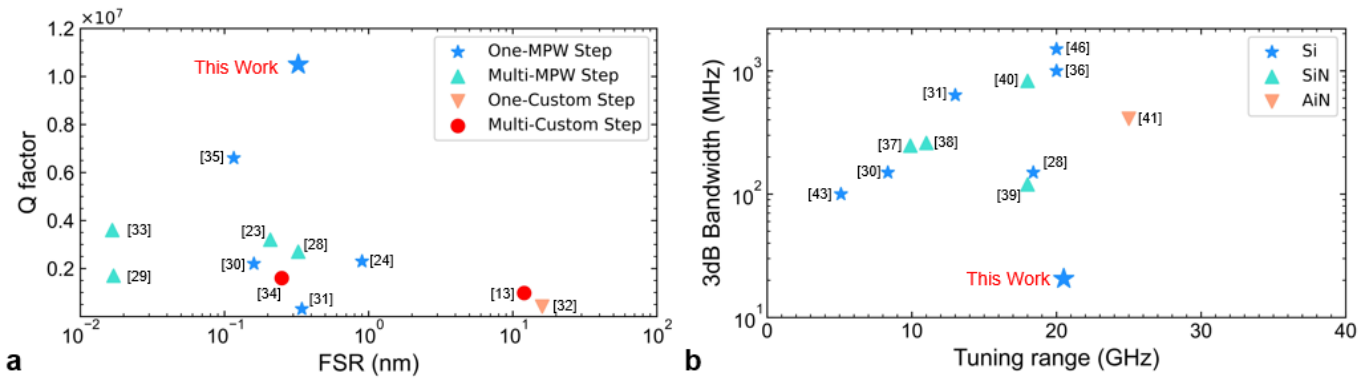

**Fig. 7.** (a) Summary of the intrinsic Q-factor as well as the FSR of the reported MRRs on a popular 220-nm-thick SOI platform; (b) Reported integrated microwave photonic filters (IMPFs) with an ultra-narrow bandwidth.

As a summary, it can be seen that these silicon photonic devices have exhibited very excellent performances enabled by broadening the core-regions to be beyond the singlemode regime, in which way the propagation loss of the fundamental mode is reduced very significantly. Meanwhile, the silicon photonic devices beyond the singlemode regime can be compact by introducing Euler MWBs. Furthermore, there is not any additional special fabrication processes. It is believed that the proposed silicon photonics beyond the singlemode regime might be the key for new-generation large-scale photonic integration and can be further readily extended to other waveguide systems.

## MARERIAL AND FABRICATION

**Material.** All the devices were designed for the transverse-electric (TE) polarization. The SOI wafer with a 220-nm-thick silicon core-layer and a 2-μm-thick buried-oxide layer. The operation wavelength is around 1550 nm, and the corresponding refractive index of silicon and silica are $n_{Si}$ = 3.455 and $n_{SiO2}$ = 1.445, respectively.

**Fabrication.** All the devices demonstrated in this article were fabricated by the MPW foundry with the standard process of deep-UV lithography technologies and inductively-coupled plasma dry-etching. A 2 μm-thick silica thin film was deposited on the top as the upper-cladding.

**Characterization.** In order to characterize the loss of the fabricated silicon photonic delayline, the setup with an ASE source and an OSA was used.

## REFERENCES

1. Thomson, D., Zilkie, A., Bowers, J. E. et al. Roadmap on silicon photonics. *J. Opt.* **18**, 073003 (2016).
2. Su, Y., Zhang, Y., Qiu, C. et al. Silicon photonic platform for passive waveguide devices: materials, fabrication, and applications. *Adv. Mater. Technol.* **5**, 1901153 (2020).
3. Bogaerts, W. & Chrostowski, L. Silicon photonics circuit design: methods, tools and challenges. *Laser Photonics Rev* **12**, 1700237 (2018).
4. Rahim, A., Spuesens, T., Baets, R. & Bogaerts, W. Open-access silicon photonics: Current status and emerging initiatives. *Proc. IEEE* **106**, 2313-2330 (2018).
5. Miller, S. E. Integrated optics: An introduction. *Bell Syst. Tech. J.* **48**, 2059-2069 (1969) .
6. Takahashi,H. Planar lightwave circuit devices for optical communication: present and future, *Proc. SPIE* **5246**, 520-531 (2003).
7. Ciminelli, C., Dell'Olio, F., Campanella, C. E. & Armenise, M. N. Photonic technologies for angular velocity sensing. *Adv. Opt. Photon.* **2**, 370-404 (2010).
8. Burmeister, E. F. et al. Photonic integrated circuit optical buffer for packet-switched networks. *Opt. Express* **17**, 6629-35 (2009).
9. Horikawa, K., Ogawa, I., Kitoh, T. & Ogawa, H. Silica-based integrated planar lightwave true-time-delay network for microwave antenna applications. In *Optical Fiber Communications*. (IEEE, 1996).
10. Henry, C. H., Kazarinov, R. F. et al. Low loss $Si_3N_4$-$SiO_2$ optical waveguides on Si. *Appl. Opt.* **26**, 2621-2624 (1987).
11. Liu, Y., Choudhary, A., Marpaung, D., & Eggleton, B. J. Integrated microwave photonic filters. *Adv. Opt. Photon.* **12**, 485-555 (2020).
12. Biberman, A., Shaw, M. J., Timurdogan, E., Wright, J. B. & Watts, M. R. Ultralow-loss silicon ring resonators. *Opt. Lett.* **37**, 4236-4238 (2012).
13. Jiang, W. C., Zhang, J., Usechak, N. G. & Lin, Q. Dispersion engineering of high-Q silicon microresonators via thermal oxidation. *Appl. Phys. Lett.* **105**, 031112 (2014).
14. Ji, X. et al. Ultra-low-loss on-chip resonators with sub-milliwatt parametric oscillation threshold. *Optica* **4**, 619-624 (2017).
15. Griffith, A., Cardenas, J., Poitras, C. B. & Lipson, M. High quality factor and high confinement silicon resonators using etchless process. *Opt. Express* **20**, 21341-21345 (2012).
16. Luo, L. W. et al. High quality factor etchless silicon photonic ring resonators. *Opt. Express* **19**, 6284-6289 (2011).
17. Jiang, W. C. et al. Compact suspended silicon microring resonators with ultrahigh quality. *Opt. Express* **22**, 1187-1192 (2014).
18. Cardenas, J., Poitras, C. B. et al. Low loss etchless silicon photonic waveguides. *Opt. Express* **17**, 4752-4757 (2009).
19. Qiu, C. et al. Fabrication, characterization and loss analysis of silicon nanowaveguides. *J. Lightwave technol.* **32**, 2303-2307 (2014).
20. Bauters, J. F., Heck, M. J. et al. Ultra-low-loss high-aspect-ratio $Si_3N_4$ waveguides. *Opt. Express* **19**, 3163-3174 (2011).
21. Dai, D. et al. Low-loss $Si_3N_4$ arrayed-waveguide grating (de) multiplexer using nano-core optical waveguides. *Opt. Express* **19**, 14130-14136 (2011).
22. Zou, Z., Zhou, L., Li, X. & Chen, J. 60-nm-thick basic photonic components and Bragg gratings on the silicon-on-insulator platform. *Opt. Express* **23**, 20784-20795 (2015).
23. Zhang, Y. et al. Design and demonstration of ultra-high-Q silicon microring resonator based on a multi-mode ridge waveguide. *Opt. Lett.* **43**, 1586-1589 (2018).
24. Li, G., Yao, J., Thacker, H. et al. Ultralow-loss, high-density SOI optical waveguide routing for macrochip interconnects. *Op. Express* **20**, 12035-12039 (2012).
25. Zhang, L. et al. Ultrahigh-Q silicon racetrack resonators. *Photonics Res.* **8**, 684-689 (2020).
26. Barwicz, T. & Haus, H. A. Three-dimensional analysis of scattering losses due to sidewall roughness in microphotonic waveguides. *J. Lightwave technol.* **23**, 2719 (2005).
27. Xu, H. et al. Silicon Integrated Nanophotonic Devices for On-Chip Multi-Mode Interconnects. *Appl. Sci.* **10**, 6365 (2020).
28. Lee, H., Chen, T., Li, J., Painter, O. & Vahala, K. J. Ultra-low-loss optical delay line on a silicon chip. *Nat. Commun.* **3**, 1-7 (2012).
29. Qiu, H., Zhou, F. et al. A continuously tunable sub-gigahertz microwave photonic bandpass filter based on an ultra-high-Q silicon microring resonator. *J. Lightwave technol.* **36**, 4312-4318 (2018).
30. Guillén-Torres, M. A. et al. Large-area, high-Q SOI ring resonators. In *IEEE Photonics Conference*. (IEEE, 2014).
31. Burla, M., Crockett, B., Chrostowski, L. & Azaña, J. Ultra-high Q multimode waveguide ring resonators for microwave photonics signal processing. In *International Topical Meeting on Microwave Photonics*. (IEEE, 2015).
32. Rasras, M. S. et al. Demonstration of a tunable microwave-photonic notch filter using low-loss silicon ring resonators. *J. Lightwave technol.* **27**, 2105-2110 (2009).
33. Xiao, S. et al. Compact silicon microring resonators with ultra-low propagation loss in the C band. *Opt. Express* **15**, 14467-14475 (2007).

34  Guillén-Torres, M. Á., Murray, K., Yun, H., Caverley, M., Cretu, E., Chrostowski, L. & Jaeger, N. A. Effects of backscattering in high-Q, large-area silicon-on-insulator ring resonators. *Opt. Lett.* **41**, 1538-1541 (2016).
35  Pafchek, R., Tummidi, R., Li, J., Webster, M. A., Chen, E. & Koch, T. L. Low-loss silicon-on-insulator shallow-ridge TE and TM waveguides formed using thermal oxidation. *Appl. Opt.* **48**, 958-963 (2009).
36  Onural, D., Gevorgyan, H., Zhang, B., Khilo, A. & Popović, M. A. Ultra-high Q resonators and sub-GHz bandwidth second order filters in an SOI foundry platform. In *Optical Fiber Communication Conference*. (Optical Society of America, 2020).
37  Sancho, J. et al. Integrable microwave filter based on a photonic crystal delay line. *Nat. Commun.* **3**, 1-9 (2012).
38  Marpaung, D. et al. $Si_3N_4$ ring resonator-based microwave photonic notch filter with an ultrahigh peak rejection. *Opt. Express* **21**, 23286-23294 (2013).
39  Zhu, Z., Liu, Y., Merklein, M., Daulay, O., Marpaung, D. & Eggleton, B. J. Positive link gain microwave photonic bandpass filter using $Si_3N_4$-ring-enabled sideband filtering and carrier suppression. *Opt. Express* **27**, 31727-31740 (2019).
40  Yang, H., Li, J., Hu, G., Yun, B. & Cui, Y. Hundred megahertz microwave photonic filter based on a high Q silicon nitride multimode microring resonator. *OSA Contin.* **3**, 1445-1455 (2020).
41  Li, J., Liu, Z., Geng, Q., Yang, S., Chen, H. & Chen, M. Method for suppressing the frequency drift of integrated microwave photonic filters. *Opt Express* **27**, 33575-33585 (2019).
42  Liu, X. et al. Broadband tunable microwave photonic phase shifter with low RF power variation in a high-Q AlN microring. *Opt. Lett.* **41**, 3599-3602 (2016).
43  Zhang, Z., Huang, B., Zhang, Z., Cheng, C. & Chen, H. Microwave photonic filter with reconfigurable and tunable bandpass response using integrated optical signal processor based on microring resonator. *Opt. Eng.* **52**, 127102 (2013).
44  Zhang, B. et al. Compact multi million Q resonators and 100 MHz passband filter bank in a thick-SOI photonics platform. *Opt. Lett.* **45**, 3005-3008 (2020).
45  Xie, Y., Choudhary, A. et al. System-level performance of chip-based Brillouin microwave photonic bandpass filters. *J. Lightwave technol.* **37**, 5246-5258 (2019).
46  Marpaung, D., Morrison, B. et al. Low-power, chip-based stimulated Brillouin scattering microwave photonic filter with ultrahigh selectivity. *Optica* **2**, 76-83 (2015).
47  Choudhary, A., Liu, Y., Morrison, B. et al. High-resolution, on-chip RF photonic signal processor using Brillouin gain shaping and RF interference. *Sci. Rep.* **7**, 1-9 (2017).
48  Wang, X. et al. Continuously tunable ultra-thin silicon waveguide optical delay line. *Optica* **4**, 507-515 (2017).
49  Dong, P., Qian, W. et al. Low loss shallow-ridge silicon waveguides. *Opt. Express* **18**, 14474-14479 (2010).
50  Bellegarde, C., Pargon, E. et al. Improvement of sidewall roughness of sub-micron silicon-on-insulator waveguides for low-loss on-chip links. In *Silicon Photonics XII*. (International Society for Optics and Photonics, 2017).
51  Takei, R., Manako, S., Omoda, E., Sakakibara, Y., Mori, M. & Kamei, T. Sub-1 dB/cm submicrometer-scale amorphous silicon waveguide for backend on-chip optical interconnect. *Opt. Express* **22**, 4779-4788 (2014).
52  Reboud, V., Blampey, B. et al. Experimental study of silicon ring resonators and ultra-low losses waveguides for efficient silicon optical interposers. In *Optical Interconnects XVI*. (International Society for Optics and Photonics, 2016).
53  Wilmart, Q., Brision, S. et al. A Complete Si Photonics Platform Embedding Ultra-Low Loss Waveguides for O-and C-Band. *J. Lightwave technol.* **39**, 532-538 (2020).
54  Cherchi, M., Harjanne, M. et al. Low-loss delay lines with small footprint on a micron-scale SOI platform. In *Silicon Photonics X*. (International Society for Optics and Photonics, 2015).

Acknowledgment

This project is supported by National Major Research and Development Program (No. 2018YFB2200200/2018YFB2200201), National Science Fund for Distinguished Young Scholars (61725503), National Natural Science Foundation of China (NSFC) (61961146003, 91950205), Zhejiang Provincial Natural Science Foundation (LZ18F050001, LD19F050001), and The Fundamental Research Funds for the Central Universities.

Author contributions

All authors contributed extensively to the work presented in this paper. L.Z., Y.W., M.Z. and D.X.D. designed the devices, L.Z., H.S.H., H.Y., T.N.C., Y.W.X., L.L. and D.X.D. performed the measurements, data analyses and discussions. H.S.H., Z.J.Y., Y.H., Y.W.X and Y.C.S. conducted theoretical analysis. L.Z., Z.J.Y. and D.X.D. wrote the manuscript. D.X.D. supervised the project.

Additional information

Supplementary information is available in the online version of the paper. Correspondence and requests for materials should be addressed to D.X.D.

Competing financial interests

The authors declare no competing financial interests.